# ROUTE TO CHAOS IN THREE-DIMENSIONAL MAPS OF LOGISTIC TYPE.

Danièle Fournier-Prunaret(*), Ricardo Lopez-Ruiz(**), Abdel-Kaddous Taha(*)[1]

**Abstract** : **A route to chaos is studied in 3-dimensional maps of logistic type. Mechanisms of period doubling for invariant closed curves (ICC) are found for specific 3-dimensional maps. These bifurcations cannot be observed for ICC in the 2-dimensional case. When the parameter of the system is modified, localized oscillations occur on the ICC that give rise to weakly chaotic rings, then to chaotic attractors, which finally disappear by contact bifurcations. These maps can be considered as models for the symbiotic interaction of three species.**

## I INTRODUCTION

Many papers have been devoted to the study of two-dimensional coupled logistic maps [1] - [4], [6]. Some of them can be considered as biological models, corresponding to interactions between species [1] - [4]. In this paper, we consider two models of the same kind in the 3-dimensional case. The first one, given by the map $T_1$, corresponds to a symbiotic interaction between correlative pairs of species :

$$T_1 \begin{cases} x_{n+1} = \lambda(3y_n+1)x_n(1-x_n), \\ y_{n+1} = \lambda(3z_n+1)y_n(1-y_n), \\ z_{n+1} = \lambda(3x_n+1)z_n(1-z_n), \end{cases} \quad (1)$$

where $\lambda$ is a real positive parameter and $(x,y,z)$ represent the species populations.

The second one, given by the map $T_2$ :

$$T_2 \begin{cases} x_{n+1} = \lambda(x_n+y_n+z_n+1)x_n(1-x_n), \\ y_{n+1} = \lambda(x_n+y_n+z_n+1)y_n(1-y_n), \\ z_{n+1} = \lambda(x_n+y_n+z_n+1)z_n(1-z_n), \end{cases} \quad (2)$$

corresponds to a global symbiotic interaction among the 3 species (see [1] for more informations on the model), $\lambda$ is also a real positive parameter and $(x,y,z)$ represent the species populations

Our aim is to study the routes to the chaos in such 3-dimensional models when the parameter $\lambda$ is modified. Then, we consider the existence of invariant closed curves and their evolutions towards chaotic attractors.

## II PERIOD DOUBLING OF INVARIANT CLOSED CURVES

Let consider first the map $T_1$. There exists a fixed point P, whose coordinates are given by :

$$\frac{1}{3}(1+\sqrt{\frac{4\lambda-3}{\lambda}})(1,1,1). \quad (3)$$

When $\lambda$ belongs to [0.75, 0.895], this point is stable ; when $\lambda$=0.895 it undergoes a Neïmark-Hopf bifurcation and gives rise to a stable invariant closed curve (ICC) $C_1$ (Figure 1a). $C_1$ undergoes a kind of period doubling (the doubling of a toric surface in 4 dimensions) when $\lambda$=0.9705 (Figure 1b-c), the ICC becoming doubled with a curle. Such phenomenon has been observed in [7][8], also for 3-dimensional maps. This kind of phenomena does not exist in 2-dimensional case. Then a second period doubling occurs when $\lambda$=0.9985, the ICC $C_1$ becomes formed with 4 curles (Figures 1d-1e). Two other ICC $C_2$ (Figures 3a-b) and $C_3$ (Figures 5a-b) exist for other parameter values and also undergo such period doubling (Figures 3c-d-e for $C_2$ and Figure 5c for $C_3$).
Concerning the map $T_2$, the same phenomena can be observed for an order 2 cyclic ICC (Figures 7a-b).

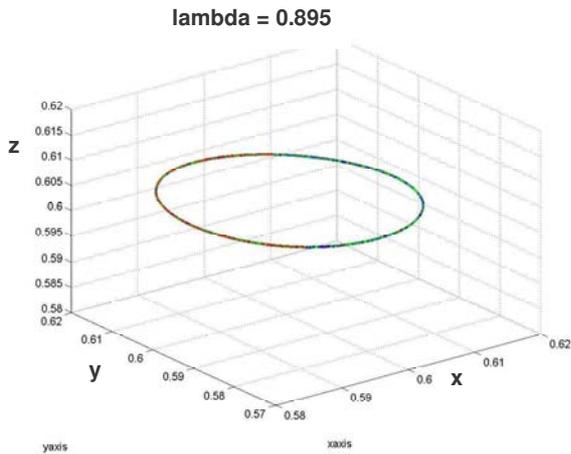

**lambda = 0.895**

Figure 1a : First invariant closed curve (ICC) $C_1$ of map $T_1$

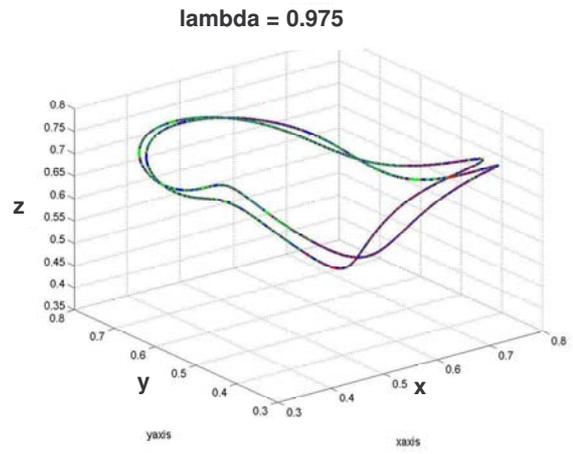

**lambda = 0.975**

Figure 1b : Period doubling of ICC $C_1$.

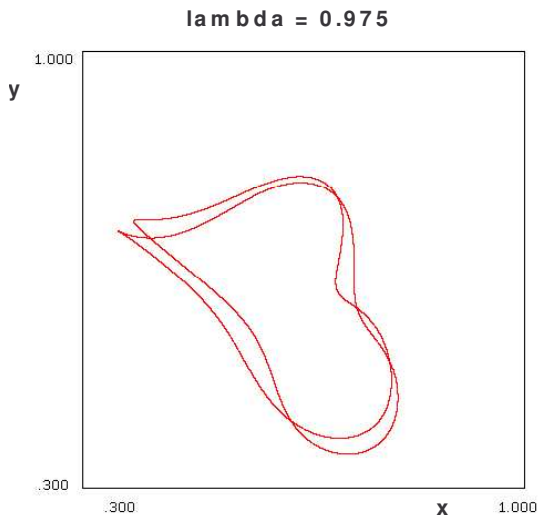

**l a m b d a = 0.975**

Figure 1c : Projection of ICC $C_1$ on $(x,y)$ plane.

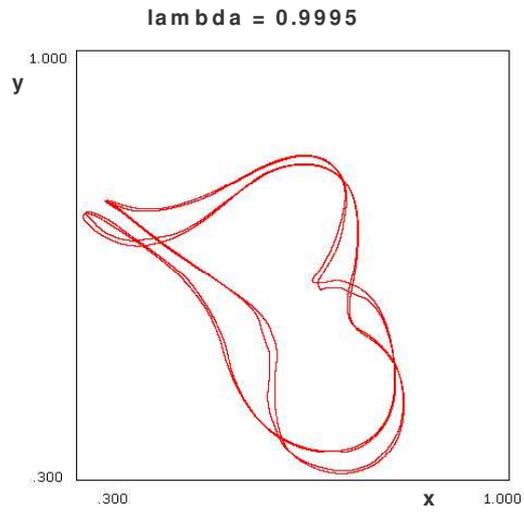

**l a m b d a = 0.9995**

Figure 1d : Projection of ICC $C_1$ on $(x,y)$ plane after second period doubling.

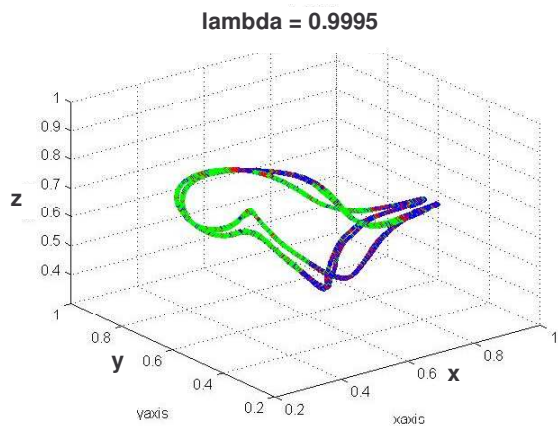

**lambda = 0.9995**

Figure 1e : ICC $C_1$ after second period doubling.

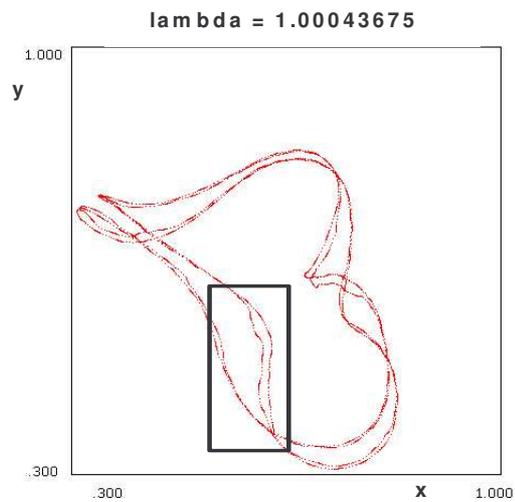

**l a m b d a = 1.00043675**

Figure 2a : Frequency locking on ICC $C_1$.

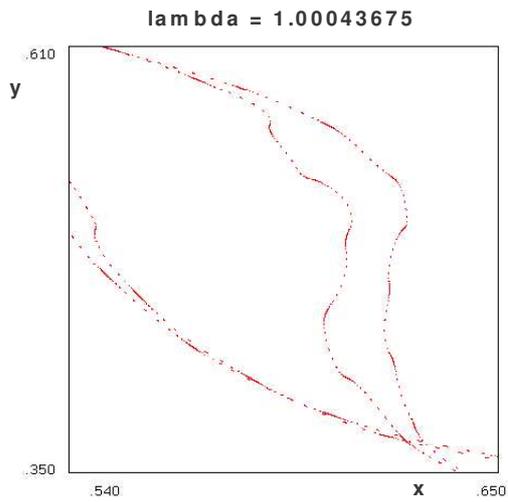

Figure 2b : Enlargment of Figure 2a.

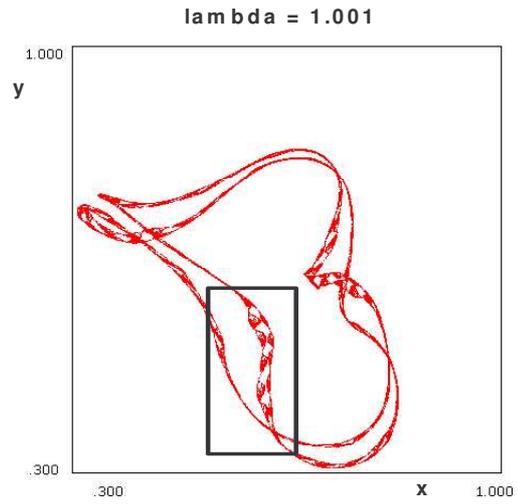

Figure 2c : ICC $C_1$ becomes a weakly chaotic ring (WCR).

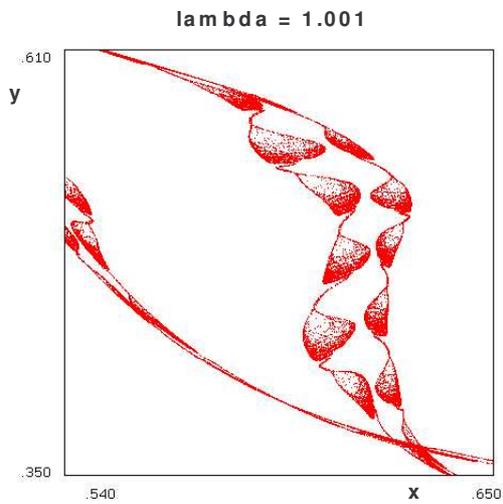

Figure 2d : Enlargment of Figure 2c.

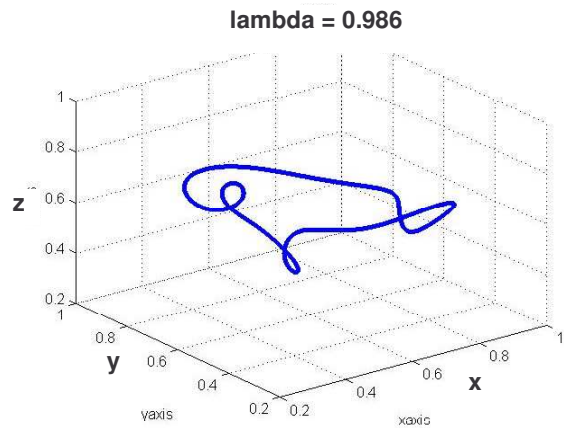

Figure 3a : Second ICC $C_2$ of map $T_1$.

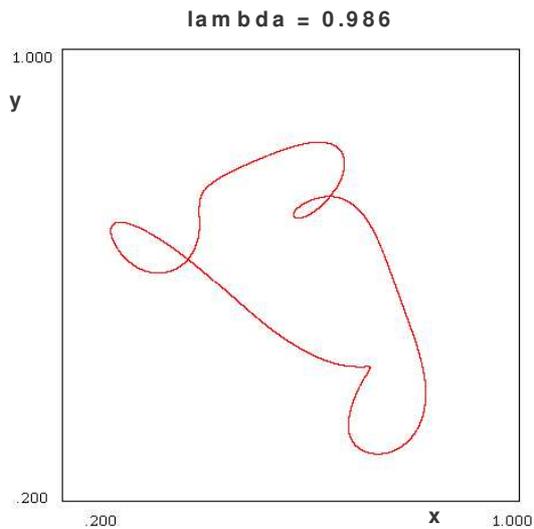

Figure 3b : Projection of ICC $C_2$ on $(x,y)$ plane.

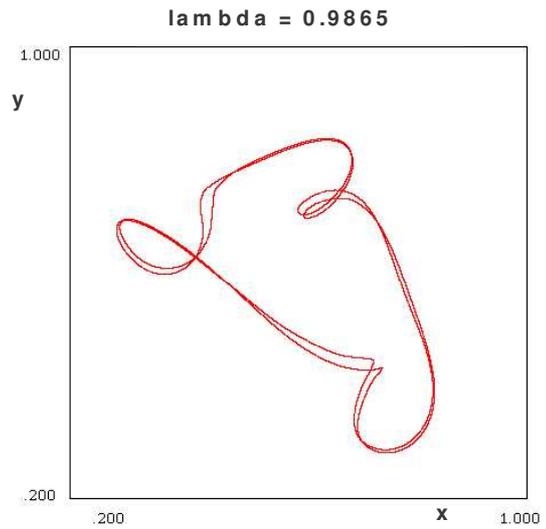

Figure 3c : Projection of ICC $C_2$ on $(x,y)$ plane after period doubling.

**lambda = 0.987**

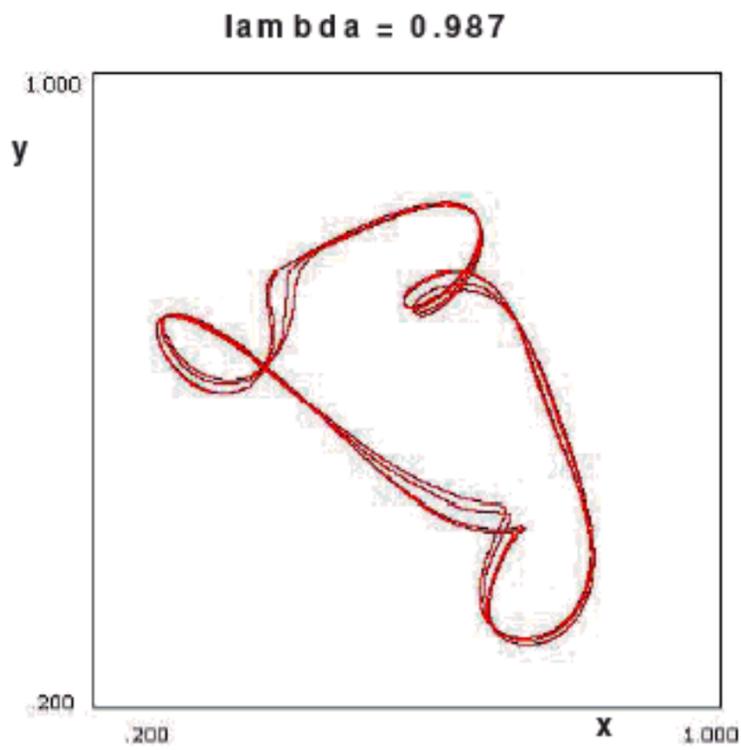

Figure 3d : Projection of ICC $C_2$ on $(x,y)$ plane after second period doubling.

**lambda = 0.987**

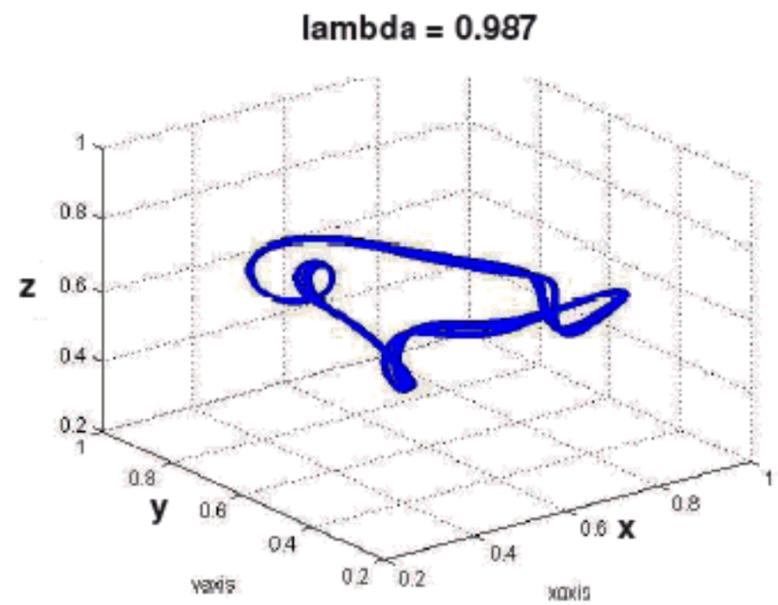

Figure 3e : ICC $C_2$ after second period doubling.

**lambda = 0.98701**

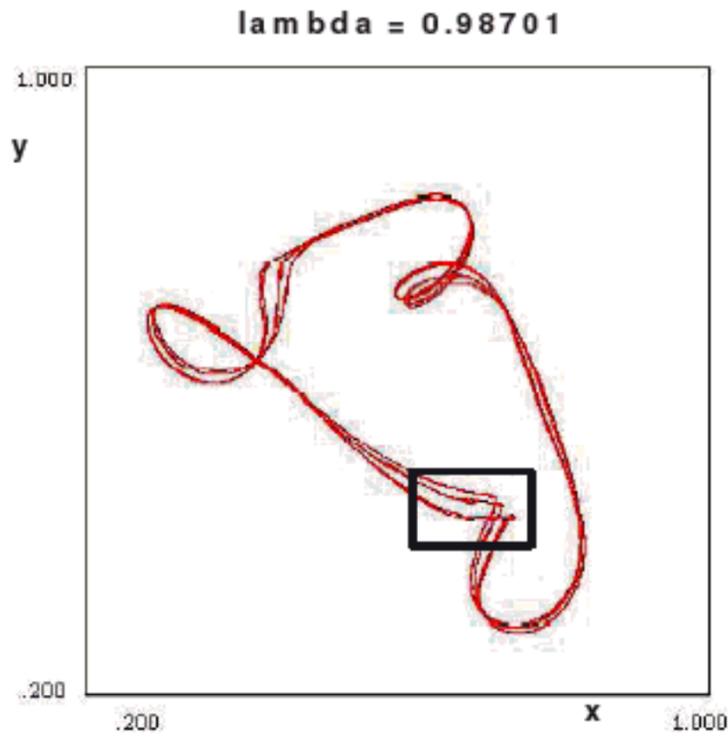

Figure 4a : Projection of ICC $C_2$ on $(x,y)$ plane after it becomes a WCR.

**lambda = 0.98701**

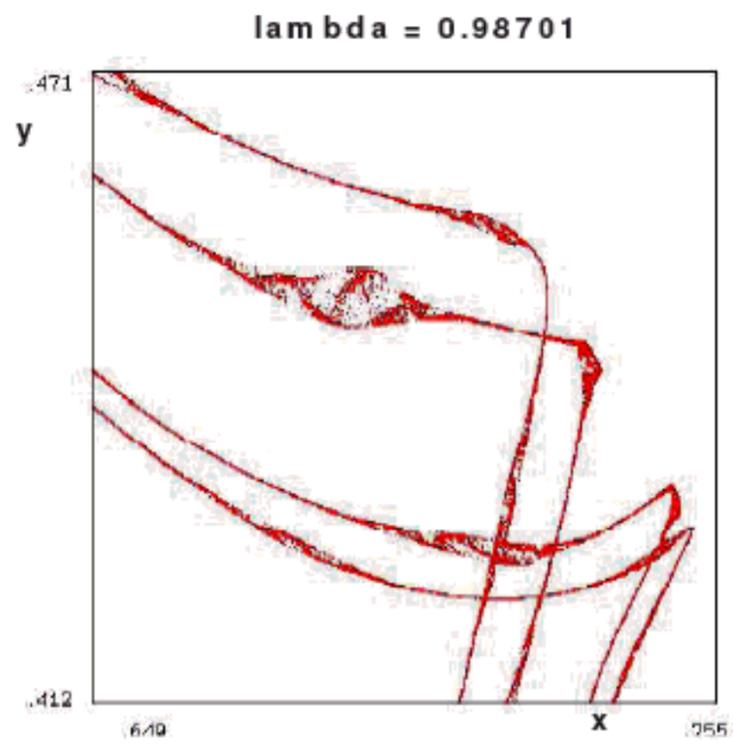

Figure 4b : Enlargment of Figure 4a.

**lambda = 1.0039**

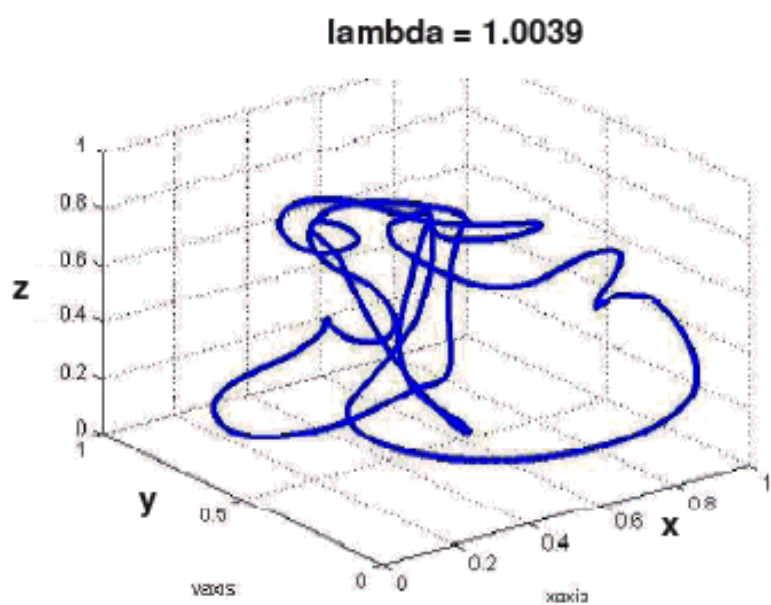

Figure 5a : Third ICC $C_3$ of map $T_1$.

**lambda = 1.0039**

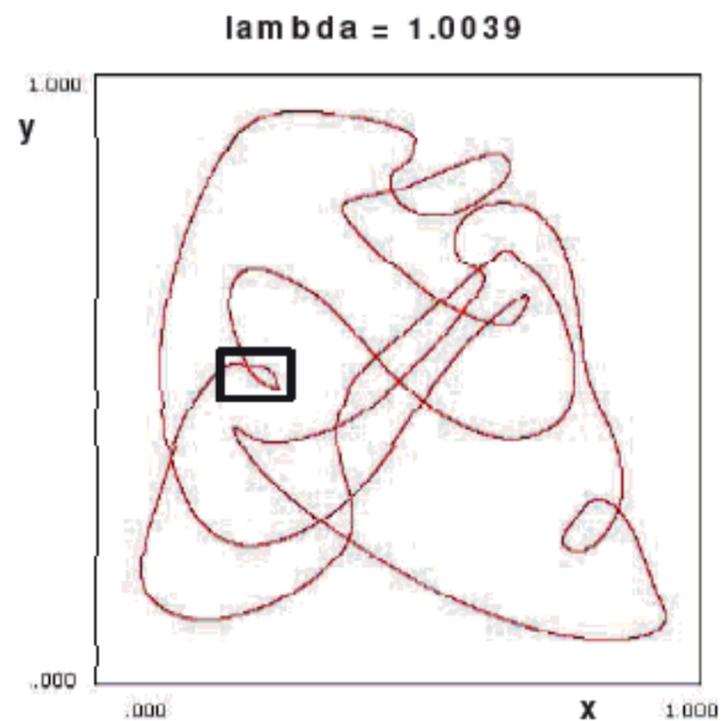

Figure 5b : Projection of ICC $C_3$ on $(x,y)$ plane.

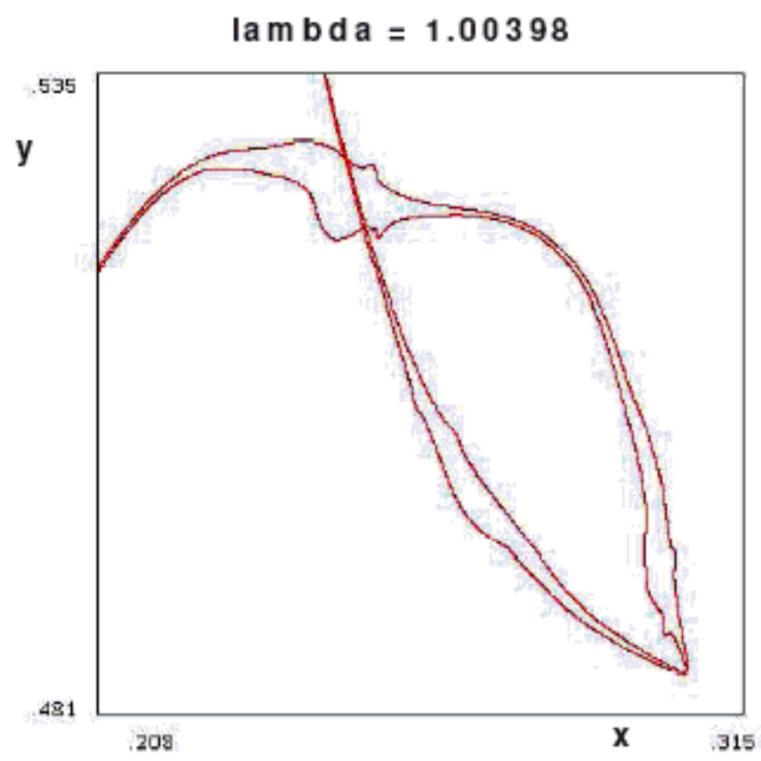

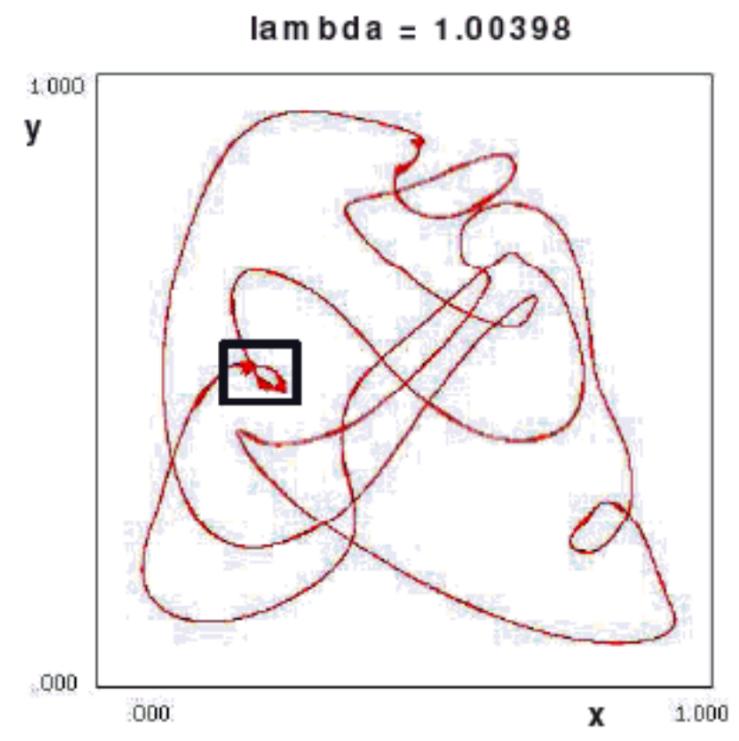

Figure 5c : Enlargment of Figure 5b, one can see a period doubling which cannot be easily seen on Figure 5b.

Figure 6a : Projection of ICC $C_3$ on $(x,y)$ plane after it becomes a WCR.

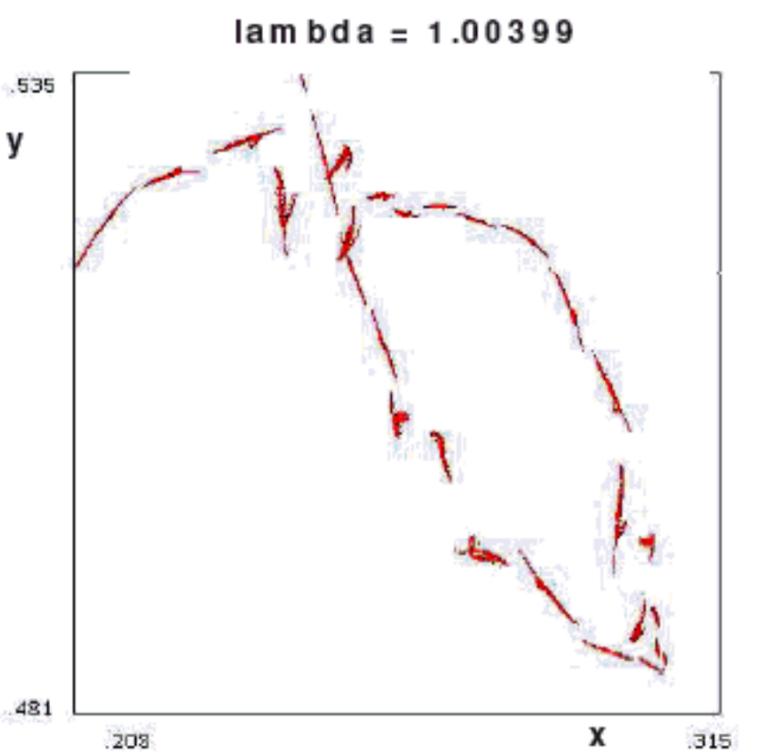

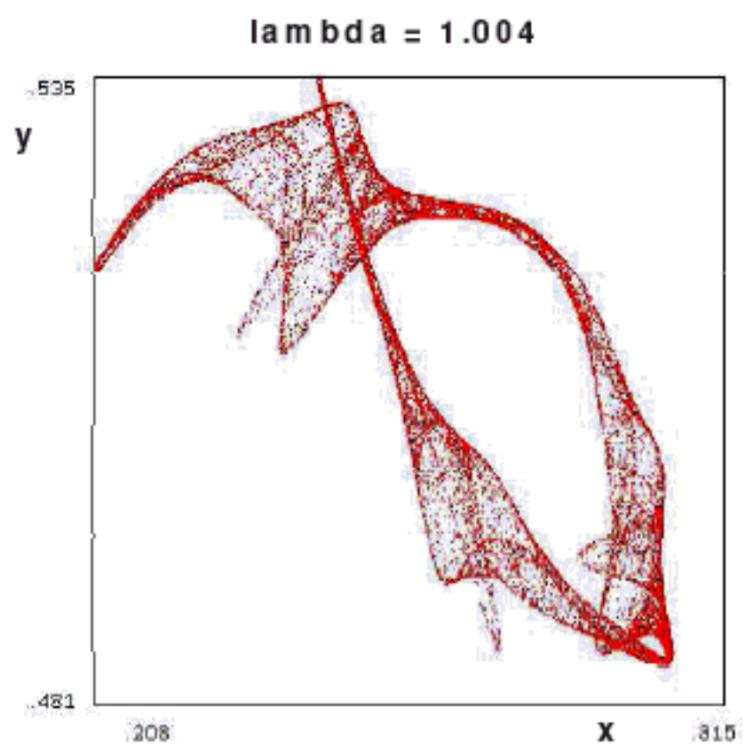

Figure 6b : Enlargment of Figure 6a.

Figure 6c : The WCR of Figure 6a becomes a chaotic attractor.

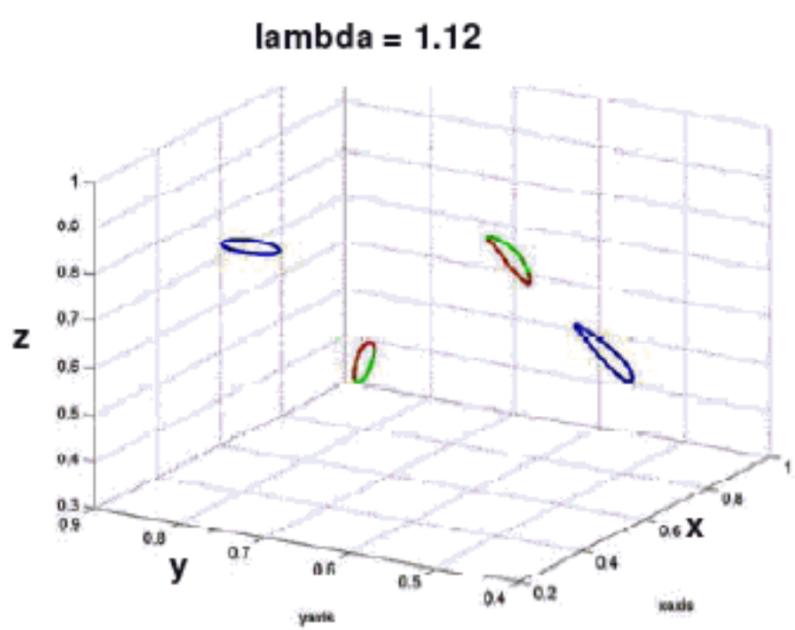

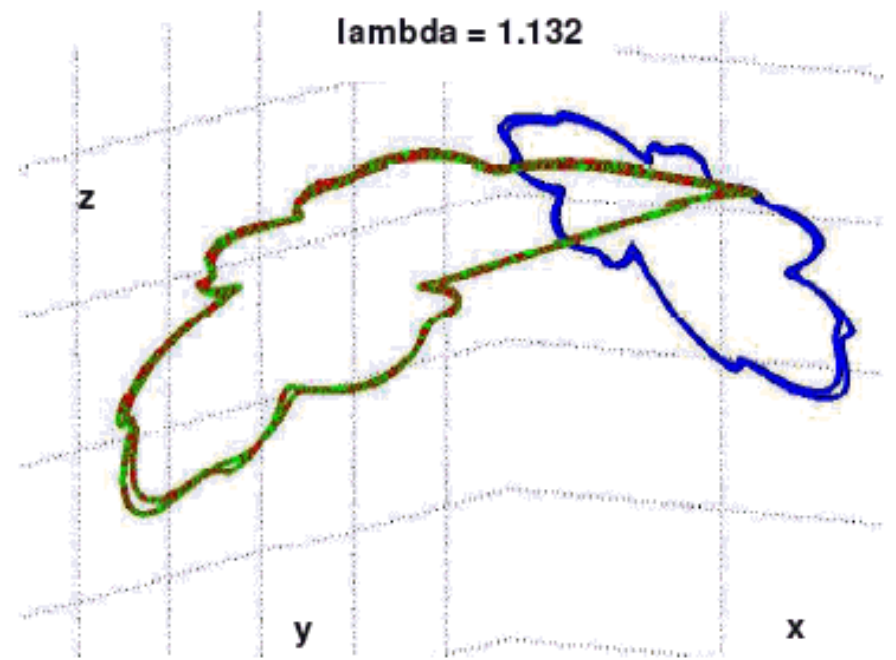

Figure 7b : Map $T_2$, period doubling of order 2 cyclic ICC.

**lambda = 1.15**

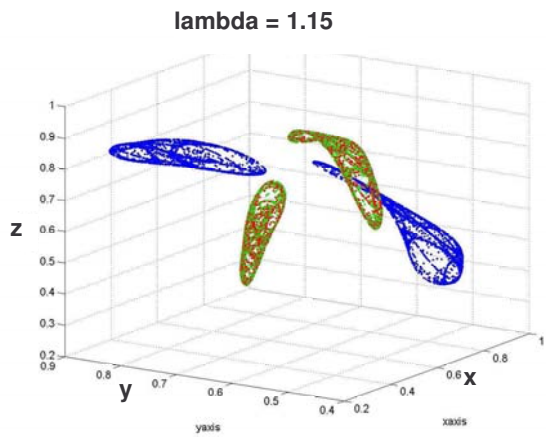

**lambda = 1.17**

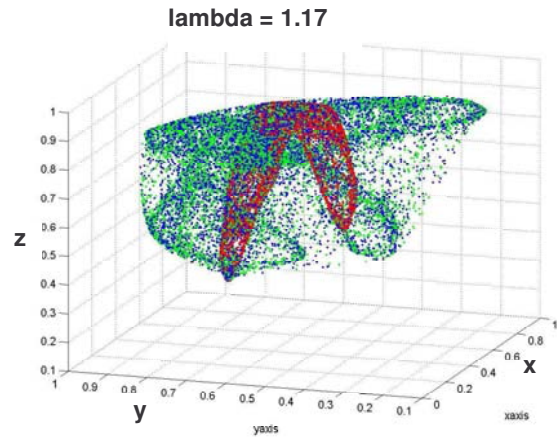

Figure 8a : The two cyclic ICC of Figures 7 become cyclic chaotic attractors.

Figure 8b : After contact bifurcation, the two cyclic chaotic attractors of Figure 8a become a single one attractor.

## III FROM WEAKLY CHAOTIC RINGS TO CHAOS

After the phenomena of period doubling for the different ICC of the two maps $T_1$ and $T_2$, bifurcations analogous to those occuring in the 2-dimensional case, can be observed. Frequency lockings (Figures 2a-b for $C_1$) and appearance of weakly chaotic rings (WCR) [5] (Figures 2c-d for ICC $C_1$, Figures 4a-b for ICC $C_2$ and Figures 6a-b for ICC $C_3$ of map $T_1$). Some "chaotic pockets" appear on parts of curves (Figures 2d & 4b) as it has also been observed for the 2-dimensional case [4].
Then WCR change to chaotic attractors (Figures 6c for ICC $C_3$ of $T_1$ and 8a for the order 2 cyclic ICC of $T_2$).

## IV MULTISTABILITY

In both cases ($T_1$ and $T_2$), phenomena of multistability can be obtained.
For $T_1$, the ICC $C_1$ coexists with an order 3 cyclic ICC (Figure 9a), which becomes an order 3 WCR (Figures 9b-c) before disappearing, as also observed in 2-dimensional case [4]. The ICC $C_1$ coexists with the ICC $C_2$ (Figure 10a) and with the chaotic attractor $C_2$ becomes (Figure 10b). For other parameter value, the ICC $C_1$ coexists with the ICC $C_3$ (Figure 11). ICC $C_2$ and $C_3$ never coexist.
For map $T_2$, the two order 2 cyclic ICC coexist before becoming two order 2 cyclic chaotic attractors, as observed in previous paragraph (Figures 7a-b & 8a). Then, by contact bifurcations of heteroclinic type, the two chaotic attractors become a single one (Figure 8b).

**lambda = 0.95**

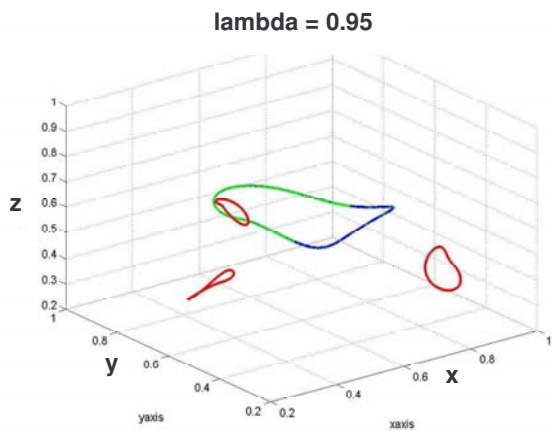

**lambda = 0.95332**

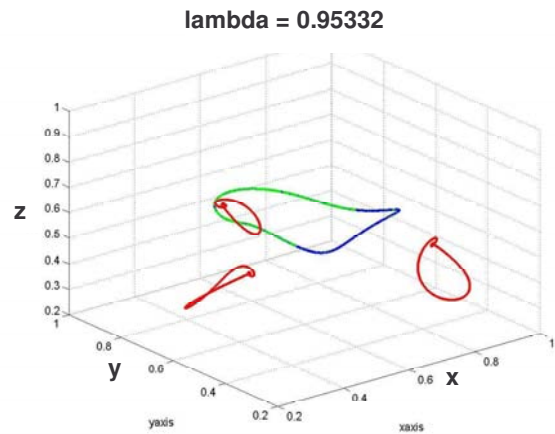

Figure 9a : Map $T_1$, coexistence of ICC $C_1$ and of an order 3 cyclic ICC.

Figure 9b : Map $T_1$, the order 3 cyclic ICC becomes a WCR.

**l a m b d a = 0.953337**

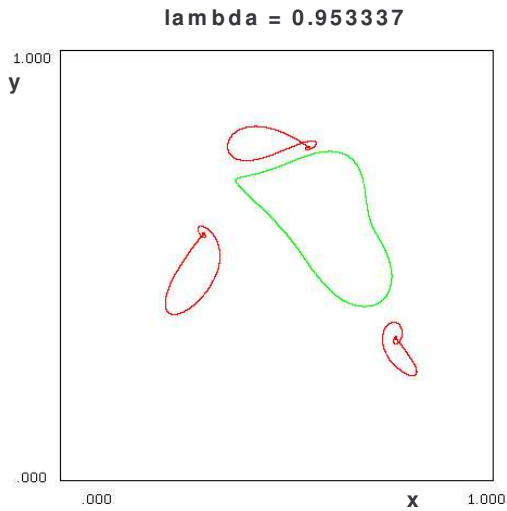

Figure 9c : Projection of ICC of Figure 9b on $(x,y)$ plane.

**lambda = 0.986**

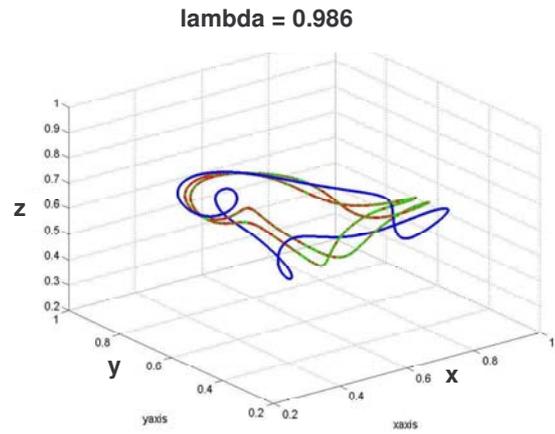

Figure 10a : Map $T_1$, coexistence of ICC $C_1$ and $C_2$.

**lambda = 0.9876**

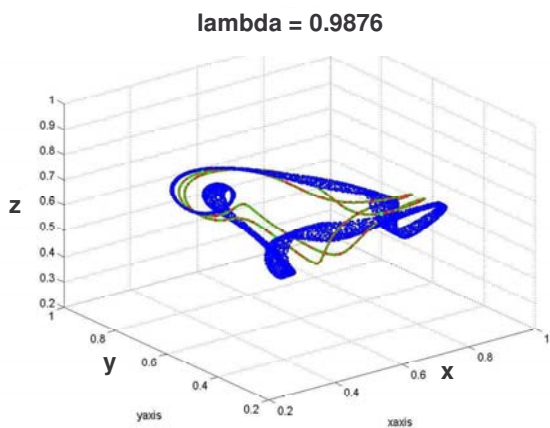

Figure 10b : Map $T_1$, coexistence of ICC $C_1$ and chaotic attractor $C_2$.

**lambda = 1.004**

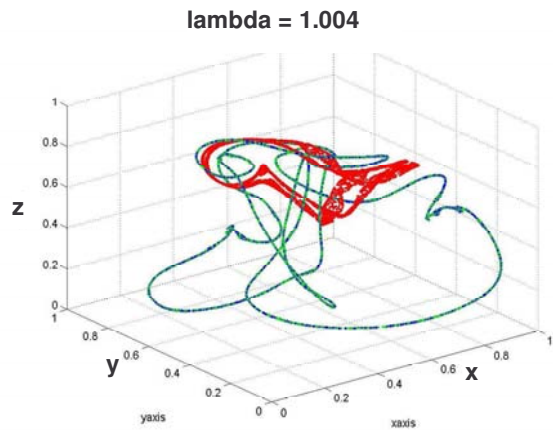

Figure 11 : Map $T_1$, coexistence of chaotic attractor $C_1$ and WCR $C_2$.

## V DISAPPEARANCE OF ATTRACTORS

As it is usual in 2-dimensional case, chaotic attractors disappear for 3-dimensional maps $T_1$ and $T_2$. One can observe two kinds of such bifurcations. The first one corresponds to the disappearance of one of both attractors by a contact bifurcation between the attractor and its basin boundary, the second attractor remains.

**lambda = 1.0040009**

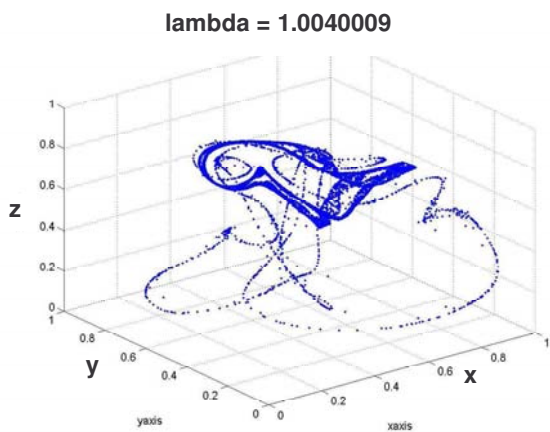

Figure 12 : Map $T_1$, the attractor $C_3$ disappears after contact bifurcation. One can see a chaotic transient close to the bifurcation before joining the attractor $C_1$.

**lambda = 0.98762**

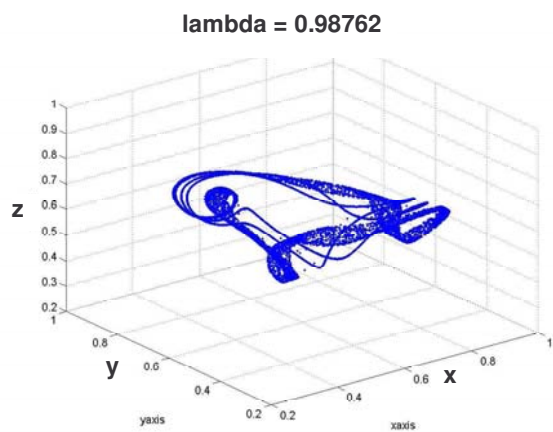

Figure 13 : Map $T_1$, the attractor $C_2$ disappears after contact bifurcation. One can see a chaotic transient close to the bifurcation before joining the attractor $C_1$.

**lambda = 1.02**

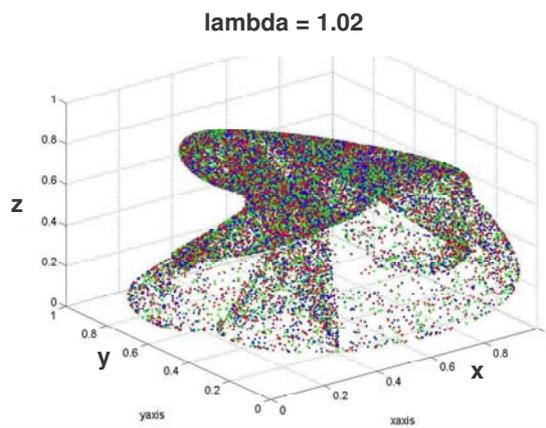

**lambda = 1.18**

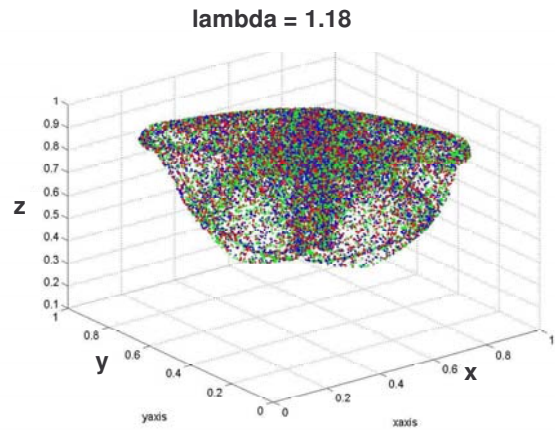

<u>Figure 14</u> : Map $T_1$, the attractor $C_1$ is going to disappear after contact bifurcation.

<u>Figure 15</u> : Map $T_2$, the chaotic attractor is going to disappear after contact bifurcation

This is the case for $C_3$ and $C_2$; the attractor $C_1$ only remains in both cases and the disappearance of $C_2$ and $C_3$ gives rise to a chaotic transient before joining the remaining attractor (Figures 12 - 13).

The second bifurcation gives rise to global unstability of the modelled system, it also corresponds to a contact bifurcation between the attractor and its basin boundary; for $T_1$, the chaotic attractor disappears when λ=1.0201 and for $T_2$, when λ=1.1805 (Figures 14-15) [5].

## VI CONCLUSION

Two 3-dimensional maps of logistic type involving a mutualistic coupling among the variables have been studied. Routes to chaos with specificities to 3-dimensional case regarding the 2-dimensional case have been observed. The most interesting phenomenon, which seems very usual in 3-dimensional case, is period doubling of ICC. Then WCR giving rise to chaotic attractors have also been observed, as contact bifurcations leading to attractor disappearance.

To complete such work, it would be of most interest to introduce studies with critical manifolds, as it usually done in the 2-dimensional case. Phenomena of multistability have been observed, so the study of basins has to be introduced, but basins are much more complicated to obtain and to analyze in the 3-dimensional case. As maps $T_1$ and $T_2$ can represent the evolution of species population, it would be also interesting to explain all this behaviour in such a context.